# Toward environmental sustainability: an empirical study on airports efficiency

Riccardo Gianluigi Serio[1], Maria Michela Dickson[2*], Diego Giuliani[2] and Giuseppe Espa[2]

[1]Doctoral School of Social Sciences, University of Trento, Italy

[2]Department of Economics and Management, University of Trento, Italy

[*]corresponding author: mariamichela.dickson@unitn.it

**Abstract.** The transition to more environmentally sustainable production processes and managerial practices is an increasingly important topic. Many industries need to undergo radical change to meet environmental sustainability requirements; the tourism industry is no exception. In this respect, a particular aspect that needs further attention is the relationship between airport performances and investments in environmental sustainability policies. This work represents a first attempt to provide empirical evidences about this relationship. Through the application of a non-parametrical method, we first assess the efficiency of the Italian airports' industry. Secondly, we investigated the relationship between airports' performance and management commitment toward the ecological transition using a Tobit regression model. The results show that airports' adherence to formal multi-year ecological transition programs has a positive and consistent impact on their performance.

# 1 Introduction

Air transport is a sector of prime importance in today's rapidly changing world. Paraphrasing the words of Bill Gates, the airplane was the first human tool to break down the walls imposed by distance, uniting people, languages, and cultures, creating the first world wide web. This idea is confirmed by the evidence that, since the birth of the first airline in the world in 1919 (KLM), this industry has experienced constant growth and evolution. Nowadays aviation contributes about 4% to the world GDP. In other words, if aviation were a country, it would occupy the seventeenth position in the world ranking for GDP [ATAG, 2020]. A milestone in the sector's evolution was the Airline Deregulation Act of 1978, which pushed towards a marked privatization process of the airports that formerly were almost entirely managed by public companies [Bailey and Baumol, 1983]. Before deregulation, airlines were not free to choose policies on fares (suggested by governments), nor to freely decide on routes and business models. This constituted an insurmountable barrier to access for potential new entrants to the market and the skies were crimped by aviation giants whose slots and related fares were not arising from a free and competitive market. The goal of the deregulation act was precisely to remove these mechanisms and set in motion the competition in the market. The airlines were now free to manage their fare policies and this led to a sudden reduction of the prices, inducing a double effect: the entry of new carriers into the market and the privatization of the airports, which until then had a role limited to provide a "core benefit" [Jarach, 2001]. The advent of private interests in airport administrations opened the gates of competition phenomena and a renewed interest in profits and efficiency.

Today, the airport network of a country is considered a strategic asset for governments. Moreover, airports are the main engine of the tourist sector and several studies have shown their significant contribution to the economic development of the regions that host them [ATAG, 2008]. The economic boost is due to the airport activities, which manifest themselves by reducing unemployment, increasing income per capita, enhancing productivity, favoring greater investment and trade as well as greater social and cultural development [Maughan et al., 2001]; [Gibbons and Wu, 2020]; [Graham, 2008]. Moreover, the performances of airports are a topic of interest for a vast array of stakeholders, including airlines, governments, passengers, and the residents of the served areas [Graham, 2008]; [Barros and Peypoch, 2008]. Airlines are strongly interested in airport benchmarking, because it can suggest to managers which airports to invest in, according to a decision-making process based on profit maximization [Barros, 2008]. It has been estimated that, without considering the revenues of the tourism sector as a whole, airports alone are able to contribute in a range between 1.4% and 2.5% to the growth of regional GDP [ACI, 2004]. For these reasons, airport efficiency is a topic of high interest in the scientific and managerial community.

Another topic of growing importance is sustainability, intended as the societal goal to reduce human impact on the environment. In particular, the attention to transition management (i.e., the set of processes through which certain aspects of society change significantly over a short time horizon) to push sustainable growth is increasing sharply [Gössling et al., 2012]. The humanity's awareness of climate change has grown in recent decades, leading governments around the world to demonstrate a renewed commitment to stemming the problem of Earth pollution (the Glasgow 2021 climate conference is a clear sign of this). The main knot lies in the fact that, although tourism gives a strong impulse to the development of the territory, this happens at the cost of a strong environmental impact. This is especially true for areas that experience a highly concentrated tourist phenomenon over a short period of time (i.e. coastal tourism in Southern Europe). The major problems are related to the excessive exploitation of land resources, consumption of water resources, coastal pollution, increased waste production, and air pollution [Weaver, 2012]; [Weaver, 2014]. Furthermore, in the air transport sector (which is undisputed leader in terms of the volume of tourists transported every year), the mentioned environmental concerns must be added to those related to aircraft movements. Noise, waste, carbon dioxide emissions, and other polluting gases released by aircraft engines into the atmosphere contribute significantly to pollution and the greenhouse effect. In addition, most of the waste generated onboard aircraft must be handled by airports [Graham, 2008]. In fact, the aviation sector accounted for 2% of global $CO_2$ emissions already in 2014, with an increase of 80% compared to 1980 and it is estimated that these are expected to increase by 45% by 2035 (European Airline Safety Association) [EASA, 2016]. Airports significantly contribute to the pollution of host regions; therefore, airports' management cannot ignore the ecological transition if it does not want to undermine future growth [Upham et al., 2003]. At the local level, on the one hand, the authorities must promote sustainable policies by encouraging companies to adopt strategies aimed at sustainability and the continuous improvement to protect the environment. On the other hand, companies can adopt voluntary certification measures, which ensure higher quality standards than those set by law (i.e., ISO 9001, ISO 14001, ISO 50001, SA 8000).

In the airport sector, Airport Carbon Accreditation (ACA) is currently the only globally institutionally recognized certification for reducing the carbon footprint. This measure was launched in 2008 by Airport Council International (ACI), through 6 certification steps: "Mapping", "Reduction", "Optimization", "Neutrality", "Transformation" and "Transition". A fundamental prerequisite for accessing this program is the accomplishment, by an accredited institution, of compliance with ISO14064 (Greenhouse Gas Accounting). By reaching the latest level of accreditation (Transition), the airport proves to be in line with the 2015 Paris Agreement, that is, to actively contribute to limiting the global average temperature rise to 1.5° C and no more than 2° C compared to pre-industrial levels.

Several works have been proposed to evaluate the performances of the airports [Barros and Dieke, 2007]. However, few have been interested in the environmental performances [Dimitriou et al., 2014], while the relationship between efficiency and environmental sustainability is still an unexplored topic. The aim of this work is to study the relationship between the performance in terms of efficiency of airports and the investments towards the adoption of more sustainable practices. To achieve this goal, a two-step analysis was conducted on a dataset of Italian airports: first, estimated the efficiency frontier through the Data Envelopment Analysis (DEA) approach. Subsequently, through a Tobit regression model, we investigated whether a relationship between efficiency and environmental sustainability exists. This was possible through the creation of a proxy variable that could assess which airports are further ahead in investments aimed at sustainability.

The paper is structured as follows. Section 2 offers a concise review of the literature. Afterward, the methods used for the analysis are presented in Section 3. Section 4 proposes a discussion of the main findings and Section 5 concludes the work.

## 2 Literature review

Efficiency is an important topic in various research streams, especially in economics and management [Chu et al., 2010]. Firms' efficiency, as well as the efficiency of the production process, are concepts widely studied in the transport sector (e.g., [Bell and Morey, 1995]), the agri-food sector, the large-scale retail trade (e.g., [Athanassopoulos and Ballantine, 1995]), telecommunications (e.g., [Collier and Storbeck, 1993]), the banking sector (e.g., [Barr and Seiford, 1994]). An important divergence in the literature on bench-marking can be traced back to the choice of empirical models used for the study. In fact, there is a stream focused on the use of parametric models, such as the stochastic frontier analysis (e.g., [Scotti et al., 2012]; [Abrate and Erbetta, 2010]), where it is necessary to establish a priori a functional form for the relationship between input, output, and inefficiency. Notwithstanding, another research flow is involved in the adoption of non-parametric models for the study of the frontier, such as Data Envelopment Analysis (DEA) modeling (e.g., [Gillen and Lall, 1997]; [Barros and Dieke, 2007]; [Gitto and Mancuso, 2012]; [Adler et al., 2013a]). DEA is a method that arises from the seminal work of Debreu (1951) and Farrell (1957), who were the first to try to measure the efficiency of a sample of production units, the so-called decision-making units (DMUs). This method, through a linear programming procedure, provides an efficiency score for each DMU with a limited number of necessary assumptions. It is based on an optimization function that defines weights to be attributed to the combination of inputs and outputs of each DMU such as to maximize the outputs (setting the level of inputs) or minimize the inputs satisfying at least a given level of output

(hereinafter output-oriented and input-oriented DEA models). Furthermore, the DEA modeling framework can be divided according to the assumptions on the returns to scale. The model assuming constant returns to scale was developed by Charnes et al. (1978), while Banker et al. (1984) added an assumption about the concavity of the frontier, allowing for variable returns to scale.

The DEA has been widely employed in the investigation of airport efficiency since the seminal work of Gillen and Lall (1997). Most of this literature, it has been devoted to the study of the phenomenon through the use of the radial DEA (which assumes proportionality in the reduction of inputs or in the increase of output), although researchers are split among those who have assumed constant returns to scale (e.g., [Sarkis and Talluri, 2004]; [Fung et al., 2008]), and those assuming variable returns to scale (e.g., [Adler and Berechman, 2001]; [Martín and Román, 2006]; [Barros and Dieke, 2007]). A relevant issue in applying DEA models concerns the size of DMUs and their comparability. In the literature has been shown that, when the DMUs have significantly different dimensions, the model with variable returns to scale could be more suitable since it does not penalize the smaller DMUs, by underestimating their performances. This also applies to DEA applications in the airport field [Liebert and Niemeier, 2013]. Some researchers, for example, Pels et al. (2003), Adler and Berechman (2001), studied the efficiency of airports through an input-oriented DEA (DEA-I), arguing that the main output, passenger traffic, was a phenomenon beyond managerial control, and therefore a difficult to maneuver lever. Others, such as Gillen and Lall (1997), Martín and Román (2006), Carlucci et al. (2018), have adopted an output-oriented DEA (DEA-O), assuming that most of the equipment for airport operation has the nature of fixed assets, therefore beyond the control of management (at least in the short run). In the DEA-I models, the goal is to reduce the quantities of inputs as much as possible while maintaining at least the current output levels. In the DEA-O models, the goal is to maximize the output levels while keeping the inputs constant and in any case at most below the current input consumption. Other authors have employed non-radial DEA (which relaxes the assumption that inputs or outputs change proportionally), through which it is possible to reach a hybrid measure of efficiency. In models such as the Slack Based Measure of efficiency (SBM) and Additive models, slacks are directly incorporated into the efficiency estimation process. For example, in the SBM model, a DMU will be considered efficient if and only if it has zero input excess and zero output shortfall.

Over time, a number of extensions to classical models have been proposed to consider several potential problems, such as, among others, the optimal number of DMUs [Bazargan and Vasigh, 2003]; [Cooper et al., 2007]; [Lam et al., 2009], the desirable number of inputs [Adler et al., 2013b], the sampling frontier [Simar and Wilson, 1998] and the change in efficiency over time [Gillen and Lall, 1997].

In the literature about the air transport sector, DEA models have been used in a two-step analysis. In the first step, the DEA is used as a tool to estimate the efficiency frontier, while in the second step regression models are used to explain the efficiency itself. An example of this is the work of Gillen and Lall (1997), who explained the efficiency of terminals and the airside with many different variables. Moreover, Sergi et al. (2020) highlighted a positive impact of the number of transits and public ownership on efficiency. The work of Carlucci et al (2018), suggests that the size, the presence of low-cost carriers, and cargo traffic, affect airports' efficiency. The intrinsic advantage of this research approach lies in the fact that the variables employed in the second step were not used to estimate the efficiency in the first one (DEA). In this way, it is possible to further discriminate the phenomena that influence the performances of the airports. Concerning the variables to be included in the models, several evidences have been highlighted in literature, such as for example the ownership of the airports [Oum et al., 2008], their sizes [Carlucci et al., 2018], the internationality of airport traffic [Chow and Fung, 2009] the location of the airports [Yoshida and Fujimoto, 2004] and their belonging to a group [Adler et al., 2013b]. However, the relationship between airport performance and environmental sustainability seems to remain particularly unexplored. An important and complete contribution to the definition of the state of the art of airport sustainability is offered by Greer et al. (2020). The authors, in addition to offering a comprehensive review of the literature, highlight the most critical areas of environmental impacts in an airport and define the short-term management levers of action to reduce pollution. Moreover, Greer et al. (2020) recognize that although interest in the topic is growing, the need for more investigation remains high.

## 3 Data and methods

The dataset used in the present study reports information on almost all Italian airports (30 units) in 2019. The sample covers over 99% of all passengers transiting through Italian airports in the year, so it reports information very close to that of the real population. The data were collected merging various sources: economic and financial information from the AIDA platform, data on passengers, movements of aircraft and goods from the Italian Statistical Institute (ISTAT) archives; data on the runways, internal equipment, and size from the National Air Transport Authority (ENAC); and, finally, some variables have been constructed by the authors, as explained in the following. Table 1 shows the data for the sample in 2019. The variable EMPLOYEES contains the total number of employees, CHINDESKS reports the number of check-in desks available in the terminal, RUNWAYMT the runway meters, PRODCOSTS the total production costs (thousands/EUR), TOTPAX (%), GOODS (%), TOTPLANES (%), contain information on the total number of airports' passengers, goods and airplanes managed as percentage of the total in Italy in the year 2019, TOTREVENUES indicates revenues totals (thousands / EUR) for the year studied, and SURFACE describes the total surface of the airport ($m^2$). It is important to note that, 6 out of 30 DMUs (denoted with * in the table) collect the information for several distinct airports, as the firms are part of a group. We decided to consider these airports as a whole because the initiatives taken by the management of a corporate group affect the management of all the airports in the group. The information about the movements of aircraft, goods, and passengers transported are reported in terms of % of the Italian total to makes it easier to compare the DMUs with each other. It can immediately be seen that more than 85% of the total goods handled in Italian airports are to be attributed to the airports of Bergamo (BGY), Milan (MXP-LIN), and Rome (CIA-FCO). At the same time, these 6 airports handle over 50% of all Italian passenger traffic.

[Table 1]

## 3.1 DEA model

As a first step, in order to estimate the performances of airports in the economics of production, we implemented a DEA model. We use both the variable returns to scale input-oriented model BCC-I [Charnes et al., 1978], and the constant returns to scale input-oriented model, CCR-I [Banker et al., 1984] with 4 inputs (EMPLOYEES, CHINDESKS, RUNWAYMT, PRODCOSTS) and 4 outputs (TOTPAX, GOODS, TOTPLANES and TOTREVENUES). To ensure the discriminatory power of the DEA in this analysis, we followed the criteria suggested in the literature on the relationship between the number of DMUs, and inputs and outputs (e.g., [Cooper et al., 2007]). Our model respects both the $N_{DMUs} > 3 \times (inputs + outputs)$ rule, and the $N_{DMUs} \geq (inputs + outputs)$ rule. The main advantage of studying efficiency through DEA models is that it is not required to define a priori the weights to be attributed to the inputs and outputs, or even a functional form. Formally, in the DEA analysis it is assumed that there are $n$ $DMU_s$ that must be evaluated and sorted in increasing order (from the worst to the best or vice versa), based on the best combination of efficiency between $m$ inputs and $s$ outputs. Thus, the $j_{th}$ production unit will consume $x_{ij}$ input units $i$ ($i = 1$ to $m$) and produce $y_{oj}$ output ($o = 1$ to $s$). In practice, there will be two vectors for the $DMU$ $j$ ($X_j, Y_j$) which respectively denote the vectors containing the observations relating to the inputs and outputs for the production unit $j$. Subsequently, the DEA method compares the $DMU_s$ to identify a set of possible linear dominant or non-dominant combinations for the $n$ $DMU_s$. The production unit $j$ will be input-dominated, if there is at least a linear combination of production units for which an input can be reduced, leaving the level of the other inputs and outputs unchanged. Similarly, the $DMU$ $j$ will be output-dominated, if there is at least a linear combination of production units for which the output is greater, leaving the level of used inputs and other outputs unchanged. This is the main distinction between the Input-oriented and Output-oriented models. As mentioned, for a long time, researchers have been uncertain about the best approach to adopt in airport benchmarking studies. For this work, we considered it appropriate to adopt the Input-oriented model, since, following the approach of Pels et al. (2003), we assume that the main outputs of the airport system are more difficult to control than the inputs for management, all other conditions being equal (e.g., it would be easier, albeit expensive, to phase out part of the check-in desks or to hire new staff, rather than increasing the number of passengers carried).

Furthermore, the wide variability (in terms of company size) within the sample suggests that returns to scale may play a role in the efficiency of $DMU_s$. In this regard, the assumption behind constant returns to scale is particularly appropriate when all $DMU_s$ are running at the best scale. In this case, the mathematical form of the problem is:

*Constant returns to scale (CCR-I)*

$$\min \theta_{\theta,\lambda},$$
$$s.t.:$$
$$\theta x_j - X\lambda \geq 0,$$
$$Y\lambda \geq y_j,$$
$$\lambda \geq 0$$

where $\theta$ is a scalar that estimates the radial contraction of all inputs, $\lambda$ is a non-negative vector of weights determined by the optimization process, and $x_j$ and $y_j$ are the input and output quantities of $DMU_j$, namely the airport under consideration. $(X)$ and $(Y)$ represent the input and output matrices respectively.

On the contrary, when the $DMU_s$ are not working at their best, the use of CCR-I specifications may result in technical efficiency measures confounded by scale efficiencies. In this case, the BCC-I model can solve the problem, adding the constraint $[\sum \lambda = 1]$. In this model, the production frontiers are spanned by the convex hull of the existing $DMU_s$. Moreover, the frontiers have piecewise linear and concave characteristics.

In addition, a comparison between the means and the standard deviations reported in Table 2 shows that the latter are significantly greater than the former. This suggests that the sample sizes are substantially non-homogeneous and hence supports the hypothesis that the DEA BCC-I model is appropriate for the Italian airport sector. Furthermore, some variables can take a null value (e.g., an airport may not manage cargo at all). This underlines the need to relax the assumption of semi-positivity of the CCR-I model, in favor of the model with variable returns. However, in order to understand how much of the overall efficiency (calculated by the CCR-I model) is due to purely technical/managerial efficiency and how much is due to scale efficiency (SE), we also implemented the input-oriented model both with constant and variable returns.

[Table 2]

## 3.2 Tobit regression

As previously mentioned, through these two DEA methods it is not possible to evaluate non-radial (non-equiproportional) contractions (expansions) of inputs (outputs), meaning that the models do not admit that an input (output) can be further reduced (increased) if one or more of the others have reached their minimum (maximum) level of contraction (expansion). Once the efficiency frontier has been estimated, the analysis progressed with a second step in which we performed a Tobit regression [Tobin, 1958] as an attempt to evaluate the relationship between efficiency and environmental sustainability. The Tobit regression model, or censored regression, is a useful tool for estimating a linear relationship when the dependent variable is simultaneously censored on the left or on the right. In practice, Tobin's model modifies the likelihood function in order to consider the non-equiprobability in sampling for each observation depending on whether the latent dependent variable has fallen above or below the threshold determined by the censorship. The general formulation of Tobin's model (1958) is:

$$y_i^* = \beta' x_i + \epsilon_i$$
$$y_i = y^*, \quad if\ 0 \leq y^* \leq 1$$
$$y_i = 0, \quad if\ y^* \leq 0$$
$$y_i = 1, \quad if\ y^* \geq 1$$
$$where$$
$$i = 1,\dots,N$$

In the case of efficiency studies, this model specification is quite common since the scores are usually between 0 and 1. The coefficients estimated by DEA for CCR-I and BCC-I are right-censored at value 1. Indeed, all the airports estimated to be efficient by the model report a score of 1. Through this method it was possible to measure the impact of environmental sustainability policies on the efficiency of Italian airports estimated through the DEA.

# 4 Results

## 4.2 Evaluation of airports performances

The first step of the analysis evaluated the efficiency performance of the $DMU_s$, in order to identify the best practice. The scores of overall technical efficiency (OTE), pure technical efficiency (PTE), scale efficiency (SE), and the returns to scale (RTS) resulting from the application of the DEA are shown in Table 3. The range of possible values for OTE, PTE and SE scores lie between 0 and 1. Values closer to 1 indicate greater efficiency, while values closer to 0 indicate that the $DMU$ is far from the efficient frontier. As for the scale efficiency, in general, it can be said that, if an airport reports SE results close to 1 (that is high scale efficiency), it should set measures to improve overall efficiency. Conversely, when the score $PTE > SE$, managers should consider the option of expanding the scale.

[Table 3]

For our models, 6 out of 30 airports are considered globally efficient (OTE= 1), while 12 out of 30 reach the purely technical efficiency frontier (PTE). The average efficiency of Italian airports is relatively high. Indeed, the average efficiency exceeds 79% considering the OTE and is even higher for the PTE (87.9%). Following these results, we can see how Italian airports are well managed on average. By comparing the scores arising from the 2 models, we can discriminate between the relative inefficiency of the $DMU_s$ due to the management of operations (PTE) and the inefficiency due to its scale (SE).

This implies that for the Alghero-Fertilia airport, for example, most of its overall inefficiency is linked to problems in operations (PTE = 0.7622), rather than to size (0.9678). In other words, Alghero airport managers should first focus on a better allocation of inputs and outputs, before considering the hypothesis of expanding the operational scale. On the other hand, considering for example the Bolzano airport, purely technical efficiency, although high, is not sufficient to guarantee a satisfactory level of overall efficiency (OTE = 0.4988). In this case, the management should consider intervening on the production scale (e.g., increasing the size), before changing the allocation of inputs and outputs. As said before, the fifth column of Table 3 shows the returns to scale estimated by the model for Italian airports. 6 out of 30 airports exhibit constant returns to scale and are efficient, suggesting that they are in their optimal production condition. However, the remaining $DMU_s$ operate under increasing returns, indicating that many airports could experience a more than proportional increase in performance from the increase in their production size. After having estimated the efficiency frontier for Italian airports, the analysis followed with the application of regression on the scores of the $DMU_s$.

## 4.2 Impact of environmental sustainability

In order to investigate the relationship between performance and environmental sustainability, we constructed a variable able to capture the commitment toward the ecological transition shown by the management. This variable (hereinafter SUSTAINABILITY), was constructed through an accurate investigation of air- ports' public disclosure (e.g., investments and actions introduced in an attempt to reduce the environmental impact of the airport), and thanks to discussions with some managers working in the air transport sector. Table 4 shows the summary statistics of variables considered in the regression model.

[Table 4]

Exploiting the 6 ACA certification steps, the variable SUSTAINABILITY was designed to assume values between 0 and 7. Level 0 includes all airports that do not currently exhibit any commitment towards sustainability that is higher than the duties established by Italian law. Level 1 includes all the companies that have publicly declared (through the website) a real commitment (e.g., reclamation investments in the area surrounding the airport, efficient water management, installation of solar panels), but which have not been admitted to in the ACA program. Scores from 2 to 7 are attributed to all companies that adhere to the ACA program (2= "Mapping", 3= "Reduction", 4= "Optimization", 5= "Neutrality", 6= "Transformation" and 7= "Transition"). In addition, we include some predictors widely employed in previous works in the literature, to ensure consistency in the model estimates. The EBITDA margin is an index widely adopted by managers to evaluate company performance which is valid also for the airport sector [Graham, 2008]. We expect that the higher the profitability, the higher the efficiency. LOWCOST reports the share (%) of passengers transported to the airport by low-cost airlines. Several researchers have evaluated the effects of the presence of low-cost carriers on airport performance: some of them report evidence that suggests an efficiency and performance improvement related to a low-cost network (e.g., [Bottasso et al., 2012]; [Cavaignac and Petiot, 2017]; [Carlucci et al., 2018]), some others found a negative relationship between the advent of low-cost carriers and airport performance (e.g., [Choo and Oum, 2013]). The effects of the ownership structure have been extensively studied, especially to assess the impacts of deregulation on the air transport industry. In general, private-owned airports might be expected to naturally be more inclined to higher efficiency levels. Some findings, however, seem to suggest the opposite (e.g., [Sergi et al., 2020]; [Barros and Dieke, 2008]; [Adler et al., 2013b]). In this paper, the ownership structure is described by the OWNERSHIP variable (0= mostly public-owned; 1= mostly private-owned). As mentioned before, some DMUs were merged to be part of the same corporate group. Therefore, we decided to check for this factor, attributing 1 to the merged DMUs and 0 to the individual airports in the GROUP variable. Membership of a group has been investigated in the literature. For example, Adler et al. (2013) found that, on average, airports be- longing to a group performed worse than independent ones. Finally, the impact of the size of the airport on its efficiency has also been extensively studied in the literature. In general, many studies have highlighted that larger airports, on average, outperform smaller ones ([Gillen and Lall, 1997]; [Pels et al., 2003]; [Coto-Millan et al., 2014]; [Carlucci et al., 2018]), while others have shown the opposite results or no association among efficiency and size [Bazargan and Vasigh, 2003]; [Abbott and Wu, 2002]. As a proxy of the airport dimensions, we computed the LOGAREAPAX variable, which is the logarithm of the area (in square meters) dedicated to passengers' hosting. Table 5 shows the results of the regression analysis on the CCR-I and BCC-I scores for Italian airports in 2019. Regression was implemented by adopting White's robust standard errors to overcome problems related to heteroskedasticity.

[Table 5]

Table 5 reports the marginal effects of each regressor on the efficiency of Italian airports. The management's adoption of policies aimed at environmental sustainability has a positive impact on the performance of the airport. Further, the coefficient shows a positive sign, and the likelihood ratio test (LT) confirms the significance of the estimate, rejecting the null hypothesis that the SUSTAINABILITY effect on the efficiency scores is equal to 0. This is an important result and, to our knowledge, the first in identifying a clear impact of sustain- able choices on airport performance. If we consider an airport's admission to the ACA program as a proxy to define the commitment of the management to pay substantial attention to the ecological transition, our results should convince the most skeptical stakeholders that environmental sustainability is not only synonymous with "green-washing" or "higher costs". Both DEA models adopted for the study are input-oriented, that is, designed to minimize inputs by satisfying at least a given level of output. The positive effect of an airport's admission to the ACA program could suggest that the improved efficiency is linked to an internal efficiency process (e.g., mapping excess $CO_2$ emissions). If we consider the impact of sustainability on purely technical efficiency (BCC-I), the effect seems to be amplified, confirming an improvement in purely operational management. At the same time, this work revealed important findings, albeit already explored in the literature. The EBITDA margin has a positive, statistically significant coefficient and with a considerable magnitude. This implies that the Italian air- ports, which transform a notable part of the sales volume into profits, are more efficient. The EBITDA margin levels can therefore be considered as an accurate, albeit synthetic and non-exhaustive measure or proxy of the performance in the airport industry. In addition, for the variable LCC (percentage of passenger traffic managed by low-cost), the model reports a positive coefficient for efficiency under the assumption of constant returns, while it is negative when assuming variable returns. This could suggest that slots dedicated to low-cost carriers contribute to boosting overall efficiency but reduce purely technical efficiency. Although our analysis does not reveal a statistical significance for this variable, this information can help airport managers to pay attention to the management of these carriers. Low-cost airlines can contribute in increasing performances and growth (scale), thanks to the network opportunities they offer, but their permanence and preponderance on traffic could undermine the operational efficiency of the airport, leading to ambiguous effects on net efficiency. The GROUP variable is also not significant, suggesting that independent airports or airports belonging to a group have, on average, similar performances. The same could be said for OWNERSHIP. Our results suggest that there is no evidence from data that private owned airports outperform public ones or vice-versa. Finally, the last control variable considered is the logarithm of the passenger area (LOGAREA- PAX). The coefficient is significant but with the opposite sign considering the overall efficiency (which seems to benefit from more spaces dedicated to passengers), and purely technical efficiency. This may suggest that space management is crucial for the airport. It is well known that an increasingly consistent source of airport profits is linked to non-aeronautical activities (e.g., shops, food & beverage). Increasing the available space can lead to the installation of more sale points and attractions for the passengers, and this can lead an airport to an over- all efficiency resulting from the generated higher revenues. However, more space can also lead to potential threats in the operational management of resources.

## 5 Conclusions

The present work constitutes an attempt to understand the relationship between airports efficiency and environmental sustainability. Specifically, the effect of airports joining programs aimed at reducing environmental impacts on overall efficiency and pure technical efficiency has been deepened through a DEA model and a Tobit regression analysis. Our results suggest that there is a statistically significant positive association between Italian airport efficiency and environmental sustainability. In fact, according to our model, airport efficiency increases as management's commitment to sustainable policies grows. In particular, we assessed sustainability by discriminating between airports that do not show any declared commitment to ecological transition practices exceeding the obligations imposed by national law, and airports that adhere to the ACA program. This constitutes an important result, as it demonstrates that the time is now ripe for managers to include environmental performance improvement policies in their strategic choices, particularly in the airport sector. Airports' managers, as well as regulatory bodies and policymakers, can take advantage of this result which suggests a relationship between sustainable investments and performance in terms of efficiency, including economic and financial efficiency. Furthermore, our study shows that financial performances also benefit from energy efficiency and green commitment. In addition to this, other considerations also emerge which are useful both to the managers working in this sector and to the legislator with a view to progress for the years to come. As the presence of partnerships with LCC increases, the complexity of management increases. However, although the network enhancement induced by the LCC supports an increase in overall efficiency, managers must pay particular attention to operational efficiency, which seems to be negatively correlated with the presence of these vectors. Moreover, the capital structure and membership of a group do not appear to significantly affect the efficiency of the airports. Finally, as the space available for passengers increases, the efficiency of the airport also increases. However, the management of spaces also deserves particular attention, as in order to contribute positively to increasing overall efficiency, management must demonstrate that it allocates spaces and resources wisely.

**Declaration of interest**

The authors declare that they have no known competing financial interests or personal relationships that could have appeared to influence the work reported in this article.

# References

[Abbott and Wu, 2002] Abbott, M. and Wu, S. (2002). Total factor productivity and efficiency of Australian airports. Australian Economic Review, 35(3):244–260.

[Abrate and Erbetta, 2010] Abrate, G. and Erbetta, F. (2010). Efficiency and patterns of service mix in airport companies: An input distance function approach. Transportation Research Part E: Logistics and Transportation Review, 46(5):693–708.

[ACI, 2004] ACI (2004). The social and economic impact of airports in Europe.

[Adler and Berechman, 2001] Adler, N. and Berechman, J. (2001). Measuring airport quality from the airlines' viewpoint: an application of data envelopment analysis. Transport Policy, 8(3):171–181.

[Adler et al., 2013a] Adler, N., Liebert, V., and Yazhemsky, E. (2013a). Benchmarking airports from a managerial perspective. Omega, 41(2):442–458.

[Adler et al., 2013b] Adler, N., Ülkü, T., and Yazhemsky, E. (2013b). Small regional airport sustainability: Lessons from benchmarking. Journal of Air Transport Management, 33:22–31.

[Andersen and Petersen, 1993] Andersen, P. and Petersen, N. C. (1993). A procedure for ranking efficient units in data envelopment analysis. Management science, 39(10):1261–1264.

[ATAG, 2008] ATAG (2008). Aviation benefits beyond borders.

[ATAG, 2020] ATAG (2020). The economic and social benefits of air transport.

[Athanassopoulos and Ballantine, 1995] Athanassopoulos, A. D. and Ballantine, J. A. (1995). Ratio and frontier analysis for assessing corporate performance: evidence from the grocery industry in the uk. Journal of the Operational Research Society, 46(4):427–440.

[Bailey and Baumol, 1983] Bailey, E. E. and Baumol, W. J. (1983). Deregulation and the theory of contestable markets. Yale J. on Reg., 1:111.

[Banker et al., 1984] Banker, R. D., Charnes, A., and Cooper, W. W. (1984). Some models for estimating technical and scale inefficiencies in data envelopment analysis. Management Science, 30(9):1078–1092.

[Banker and Natarajan, 2008] Banker, R. D. and Natarajan, R. (2008). Evaluating contextual variables affecting productivity using data envelopment analysis. Operations research, 56(1):48–58.

[Barr and Seiford, 1994] Barr, R. and Seiford, L. (1994). Benchmarking with data envelopment analysis. In ORSA/TIMS Joint National Meeting, Detroit, MI, volume 23.

[Barros, 2008] Barros, C. P. (2008). Airports in Argentina: Technical efficiency in the context of an economic crisis. Journal of Air Transport Management, 14(6):315–319.

[Barros and Dieke, 2007] Barros, C. P. and Dieke, P. U. (2007). Performance evaluation of Italian airports: A data envelopment analysis. Journal of Air Transport Management, 13(4):184–191.

[Barros and Dieke, 2008] Barros, C. P. and Dieke, P. U. (2008). Measuring the economic efficiency of airports: A Simar–Wilson methodology analysis. Transportation Research Part E: Logistics and Transportation Review, 44(6):1039–1051.

[Barros and Peypoch, 2008] Barros, C. P. and Peypoch, N. (2008). A comparative analysis of productivity change in Italian and Portuguese airports. International Journal of Transport Economics, 35(2):205–216. Cited By :11.

[Barros and Weber, 2009] Barros, C. P. and Weber, W. L. (2009). Productivity growth and biased technological change in UK airports. Transportation Research Part E: Logistics and Transportation Review, 45(4):642–653.

[Bazargan and Vasigh, 2003] Bazargan, M. and Vasigh, B. (2003). Size versus effi- ciency: a case study of us commercial airports. Journal of air transport management, 9(3):187–193.

[Bell and Morey, 1995] Bell, R. A. and Morey, R. C. (1995). Increasing the efficiency of corporate travel management through macro benchmarking. Journal of Travel Research, 33(3):11–20.

[Bojnec and Papler, 2011] Bojnec, S. and Papler, D. (2011). Economic efficiency, energy consumption and sustainable development. Journal of Business Economics and Management, 12(2):353–374.

[Bottasso et al., 2012] Bottasso, A., Conti, M., and Piga, C. (2012). Low-cost carriers and airports' performance: empirical evidence from a panel of UK airports. Industrial and Corporate Change, 22(3):745–769.

[Carlucci et al., 2018] Carlucci, F., Cirà, A., and Coccorese, P. (2018). Measuring and explaining airport efficiency and sustainability: Evidence from Italy. Sustainability, 10(2):400.

[Cavaignac and Petiot, 2017] Cavaignac, L. and Petiot, R. (2017). A quarter century of data envelopment analysis applied to the transport sector: A bibliometric analysis. Socio-Economic Planning Sciences, 57:84–96.

[Caves et al., 1982] Caves, D. W., Christensen, L. R., and Diewert, W. E. (1982). The economic theory of index numbers and the measurement of input, output, and productivity. Econometrica: Journal of the Econometric Society, pages 1393–1414.

[Charnes et al., 1978] Charnes, A., Cooper, W. W., and Rhodes, E. (1978). Measuring the efficiency of decision-making units. European journal of operational research, 2(6):429–444.

[Choo and Oum, 2013] Choo, Y. Y. and Oum, T. H. (2013). Impacts of low-cost carrier services on efficiency of the major US airports. Journal of Air Transport Management, 33:60–67.

[Chow and Fung, 2009] Chow, C. K. W. and Fung, M. K. Y. (2009). Efficiencies and scope economies of Chinese airports in moving passengers and cargo. Journal of Air Transport Management, 15(6):324–329.

[Chu et al., 2010] Chu, Y., Yu, J., and bin Huang, Y. (2010). Measuring airport production efficiency based on two-stage correlative DEA. In 2010 IEEE 17Th International Conference on Industrial Engineering and Engineering Management. IEEE.

[Collier and Storbeck, 1993] Collier, D. A. and Storbeck, J. E. (1993). A data envelopment approach to benchmarking in the telecommunications industry. Ohio State Faculty of Management Science Working Paper, Ohio State University, Columbus.

[Cooper et al., 2007] Cooper, W. W., Seiford, L. M., and Tone, K. (2007). Data envelopment analysis: a comprehensive text with models, applications, references and DEA-solver software, volume 2. Springer.

[Coto-Millán et al., 2014] Coto-Millán, P., Casares-Hontañón, P., Inglada, V., Agüeros, M., Pesquera, M. A., and Badiola, A. (2014). Small is beautiful? The impact of economic crisis, low cost carriers, and size on efficiency in Spanish airports (2009–2011). Journal of Air Transport Management, 40:34–41.

[Debreu, 1951] Debreu, G. (1951). The coefficient of resource utilization. Econometrica, 19(3):273.

[Dimitriou, 2018] Dimitriou, D. (2018). Air transport economic footprint in remote tourist regions. Mobilities, Tourism and Travel Behavior: Contexts and Boundaries, page 143.

[Dimitriou et al., 2014] Dimitriou, D., Voskaki, A., and Sartzetaki, M. (2014). Airports environmental management: Results from the evaluation of European airports environmental plans. International Journal of Information Systems and Supply Chain Management (IJISSCM), 7(1):1–14.

[EASA, 2016] EASA (2016). European aviation environmental report 2016.

[Farrell, 1957] Farrell, M. J. (1957). The measurement of productive efficiency. Journal of the Royal Statistical Society. Series A (General), 120(3):253.

[Fung et al., 2008] Fung, M. K. Y., Wan, K. K. H., Hui, Y. V., and Law, J. S. (2008). Productivity changes in Chinese airports 1995–2004. Transportation Research Part E: Logistics and Transportation Review, 44(3):521–542.

[Gibbons and Wu, 2020] Gibbons, S. and Wu, W. (2020). Airports, access and local economic performance: evidence from china. Journal of Economic Geography, 20(4):903–937.

[Gillen and Lall, 1997] Gillen, D. and Lall, A. (1997). Developing measures of airport productivity and performance: an application of data envelopment analysis. Transportation Research Part E: Logistics and Transportation Review, 33(4):261–273.

[Gitto and Mancuso, 2012] Gitto, S. and Mancuso, P. (2012). Two faces of airport business: A non-parametric analysis of the Italian airport industry. Journal of Air Transport Management, 20:39–42.

[Gossling et al., 2012] Gossling, S., Hall, C. M., Ekström, F., Engeset, A. B., and Aall, C. (2012). Transition management: a tool for implementing sustainable tourism scenarios? Journal of Sustainable Tourism, 20(6):899–916.

[Graham, 2008] Graham, A. (2008). Managing airports: An international perspective (3rd ed.). Elsevier.

[Greer et al., 2020] Greer, F., Rakas, J., and Horvath, A. (2020). Airports and environmental sustainability: a comprehensive review. Environmental Research Letters, 15(10).

[Jarach, 2001] Jarach, D. (2001). The evolution of airport management practices: to- wards a multi-point, multi-service, marketing-driven firm. Journal of air transport management, 7(2):119–125.

[Lam et al., 2009] Lam, S. W., Low, J. M., and Tang, L. C. (2009). Operational efficiencies across Asia pacific airports. Transportation Research Part E: Logistics and Transportation Review, 45(4):654–665.

[Liebert and Niemeier, 2013] Liebert, V. and Niemeier, H.-M. (2013). A survey of empirical research on the productivity and efficiency measurement of airports. Journal of Transport Economics and Policy (JTEP), 47(2):157–189.

[Martín and Roman, 2006] Martín, J. C. and Roman, C. (2006). A benchmarking analysis of Spanish commercial airports. a comparison between smop and dea ranking methods. Networks and Spatial Economics, 6(2):111–134.

[Martín and Roman, 2001] Martín, J. C. and Roman, C. (2001). An application of DEA to measure the efficiency of spanish airports prior to privatization. Journal of Air Transport Management, 7(3):149–157.

[Maughan et al., 2001] Maughan, J., Raper, D., Thomas, C., and Gillingwater, D. (2001). Scan-uk the uk sustainable cities and aviation network. Air & Space Europe, 3(1-2):56–59.

[Murillo-Melchor, 1999] Murillo-Melchor, C. (1999). An analysis of technical efficiency and productivity changes in Spanish airports using the Malmquist index. International Journal of Transport Economics/Rivista internazionale di economia dei trasporti, pp. 271–292.

[Oum et al., 2008] Oum, T. H., Yan, J., and Yu, C. (2008). Ownership forms matter for airport efficiency: A stochastic frontier investigation of worldwide airports. Journal of urban economics, 64(2):422–435.

[Pels et al., 2003] Pels, E., Nijkamp, P., and Rietveld, P. (2003). Inefficiencies and scale economies of european airport operations. Transportation Research Part E: Logistics and Transportation Review, 39(5):341–361.

[Perrakis, 1982] Perrakis, S. (1982). Contestable markets and the theory of industry structure.

[Sarkis and Talluri, 2004] Sarkis, J. and Talluri, S. (2004). Performance based cluster- ing for benchmarking of US airports. Transportation Research Part A: Policy and Practice, 38(5):329–346.

[Scotti et al., 2012] Scotti, D., Malighetti, P., Martini, G., and Volta, N. (2012). The impact of airport competition on technical efficiency: A stochastic frontier analysis applied to Italian airport. Journal of Air Transport Management, 22:9–15.

[Sergi et al., 2020] Sergi, S. B., D'Aleo, V., Arbolino, R., Carlucci, F., Barilla, D., and Ioppolo, G. (2020). Evaluation of the Italian transport infrastructures: A technical and economic efficiency analysis. Land use policy, 99:104961.

[Simar and Wilson, 1998] Simar, L. and Wilson, P. W. (1998). Sensitivity analysis of efficiency scores: How to bootstrap in nonparametric frontier models. Management science, 44(1):49–61.

[Simar and Wilson, 2007] Simar, L. and Wilson, P. W. (2007). Estimation and inference in two-stage, semi-parametric models of production processes. Journal of econometrics, 136(1):31–64.

[Tobin, 1958] Tobin, J. (1958). Estimation of relationships for limited dependent variables. Econometrica: journal of the Econometric Society, pages 24–36.

[Upham et al., 2003] Upham, P., Thomas, C., Gillingwater, D., and Raper, D. (2003). Environmental capacity and airport operations: current issues and future prospects. Journal of Air Transport Management, 9(3):145–151.

[Weaver, 2012] Weaver, D. B. (2012). Organic, incremental and induced paths to sustainable mass tourism convergence. Tourism Management, 33(5):1030–1037.

[Weaver, 2014] Weaver, D. B. (2014). Asymmetrical dialectics of sustainable tourism: Toward enlightened mass tourism. Journal of Travel Research, 53(2):131–140.

[Yokomi, 2005] Yokomi, M. (2005). Measurement of Malmquist index of privatized baa plc. In 9th ATRS Conference.

[Yoshida and Fujimoto, 2004] Yoshida, Y. and Fujimoto, H. (2004). Japanese-airport benchmarking with the DEA and endogenous-weight tfp methods: testing the criticism of overinvestment in Japanese regional airports. Transportation Research Part E: Logistics and Transportation Review, 40(6):533–546.

| | AIRPORTS | IATA CODE | EMPLOYEES | CHINDESKS | RUNWAYMT | PRODCOSTS | TOTPAX (%) | GOODS (%) | TOTPLANES (%) | TOTREVENUES | SURFACE |
|---|---|---|---|---|---|---|---|---|---|---|---|
| 1 | Alghero-Fertilia | AHO | 206 | 17 | 3000 | 16831.127 | 0.72 | 0.0001 | 0.6987 | 18469.917 | 17000 |
| 2 | Ancona-Falconara | AOI | 81 | 12 | 2991 | 14504.889 | 0.2504 | 0.6582 | 0.3991 | 7305.324 | 15450 |
| 3 | * Bari-Brindisi-Foggia-Taranto | BRI-BDS-FOG-TAR | 334 | 46 | 10408 | 97032.699 | 4.3116 | 1.0228 | 3.97 | 105012.485 | 49800 |
| 4 | Bergamo-Orio Al Serio | BGY | 229 | 45 | 3024 | 120833.675 | 7.1958 | 11.286 | 6.3018 | 141991.349 | 35000 |
| 5 | Bologna-Borgo Panigale | BLQ | 519 | 74 | 2880 | 89300 | 4.9173 | 3.6062 | 5.0179 | 119180 | 44000 |
| 6 | Bolzano | BZO | 25 | 3 | 1275 | 5410.697 | 0.0006 | 0 | 0.0027 | 4331.885 | 800 |
| 7 | Cagliari-Elmas | CAG | 152 | 42 | 2805 | 49607.435 | 2.4845 | 0.4035 | 2.3831 | 54512.496 | 41290 |
| 8 | Catania-Fontanarossa | CTA | 164 | 46 | 2438 | 69424.486 | 5.3022 | 0.5473 | 5.03 | 88966.239 | 43110 |
| 9 | Cuneo-Levaldigi | CUF | 25 | 6 | 2104 | 3774.196 | 0.0467 | 0 | 0.0425 | 3034.098 | 4350 |
| 10 | * Firenze-Pisa | FLR-PSA | 336 | 56 | 4752 | 78678 | 4.2795 | 1.2043 | 4.5903 | 99416 | 56550 |
| 11 | Genova-Sestri | GOA | 203 | 11 | 3065 | 31435.988 | 0.8001 | 0.0136 | 1.0279 | 33104.356 | 12550 |
| 12 | * LameziaTerme-ReggioCalabria-Crotone | SUF-REG-CRV | 123 | 33 | 6396 | 27281.702 | 1.8378 | 0.1194 | 1.7813 | 29632.95 | 25650 |
| 13 | * Milano-Linate-Malpensa | LIN-MXP | 2731 | 293 | 10282 | 568600 | 18.3179 | 53.6279 | 20.5384 | 736699 | 364765 |
| 14 | Napoli-Capodichino | NAP | 509 | 55 | 2640 | 110375.234 | 5.6315 | 0.9419 | 5.3884 | 150788.081 | 30700 |
| 15 | Olbia-Costa Smeralda | OLB | 273 | 36 | 2446 | 34824.882 | 1.5237 | 0.0086 | 1.5346 | 55099.748 | 43800 |
| 16 | Palermo-Punta Raisi | PMO | 266 | 28 | 3326 | 70545.618 | 3.6678 | 0.1108 | 3.5852 | 78751.919 | 35400 |
| 17 | Parma | PMF | 33 | 9 | 2300 | 5159.244 | 0.0376 | 0 | 0.0355 | 1908.23 | 3700 |
| 18 | Perugia | PEG | 36 | 2 | 2199 | 5004.722 | 0.1134 | 0 | 0.1026 | 4853.245 | 1150 |
| 19 | Pescara | PSR | 38 | 8 | 2430 | 9113.769 | 0.3635 | 0.0253 | 0.3482 | 9163.377 | 11150 |
| 20 | Rimini-Miramare | RMI | 20 | 7 | 2995 | 6379.335 | 0.205 | 0.0004 | 0.1768 | 7541.669 | 15500 |
| 21 | * Roma-Ciampino-Fiumicino | CIA-FCO | 1401 | 461 | 16909 | 703507 | 25.5993 | 20.1545 | 23.7787 | 1109272 | 339150 |
| 22 | Torino-Caselle | TRN | 233 | 38 | 3300 | 53765.124 | 2.0616 | 0.0193 | 2.381 | 67133.138 | 51150 |
| 23 | Trapani-Birgi | TPS | 74 | 15 | 2690 | 8873.848 | 0.2113 | 0.0011 | 0.3097 | 4628.598 | 14700 |
| 24 | Treviso-Sant'Angelo | TSF | 165 | 16 | 2459 | 27540.456 | 1.688 | 0 | 1.3079 | 29301.408 | 11500 |
| 25 | Trieste-Ronchi dei Legionari | TRS | 108 | 12 | 3000 | 13982.967 | 0.4053 | 0.0064 | 0.5843 | 17185.147 | 23565 |
| 26 | Venezia-Tessera | VIC | 470 | 60 | 6080 | 126366 | 5.999 | 4.9311 | 6.1286 | 202848 | 53000 |
| 27 | * Verona-Brescia | VRN-VBS | 129 | 48 | 6058 | 41847.787 | 1.8842 | 1.3095 | 2.2578 | 46947.79 | 29000 |
| 28 | Lampedusa | LMP | 44 | 4 | 1800 | 4818.552 | 0.1434 | 0.0017 | 0.2883 | 4204.858 | 1300 |
| 29 | Elba | EBA | 10 | 4 | 1095 | 1082.177 | 0.001 | 0 | 0.0082 | 1117.699 | 475 |
| 30 | Grosseto | GRS | 5 | 2 | 3007 | 760.192 | 0.0001 | 0 | 0.0004 | 788.357 | 1400 |

Table 1: Data about Italian airports. Units denoted with * are the result of corporate groups.

|  | Min | 1st Q | Median | Mean | 3rd Q | Max | St.Dev |
|---|---|---|---|---|---|---|---|
| EMPLOYEES | 5.0 | 39.5 | 158.0 | 298.1 | 271.2 | 2731.0 | 531.9 |
| CHINDESKS | 2.0 | 8.2 | 22.5 | 49.6 | 46.0 | 461.0 | 94.0 |
| RUNWAYMT | 1095.0 | 2440.0 | 2993.0 | 4005.1 | 3319.5 | 16909.0 | 3302.0 |
| PRODCOSTS | 760.2 | 7003.0 | 29488.2 | 79888.7 | 76644.9 | 703507.0 | 156987.7 |
| TOTPAX | 247.0 | 397486.5 | 3089526.0 | 6413016.2 | 8279738.5 | 49250548.0 | 10740580.8 |
| GOODS | 0.0 | 1.8 | 235.5 | 35158.9 | 10574.8 | 565649.0 | 109396.3 |
| TOTPLANES | 6.0 | 4271.8 | 20673.0 | 48484.8 | 64512.2 | 345872.0 | 80419.2 |
| TOTREVENUES | 788.4 | 5466.3 | 31368.7 | 107773.0 | 96803.6 | 1109272.0 | 233009.9 |

Table 2: Summary statistics of the variables about Italian airports.

| N. Airports | $OTE_{crs}$ | $PTE_{vrs}$ | SE | RTS |
|---|---|---|---|---|
| 1 Bergamo-Orio Al Serio | 1.00 | 1.00 | 1.00 | Constant |
| 2 Catania-Fontanarossa | 1.00 | 1.00 | 1.00 | Constant |
| 3 Milano-Linate-Malpensa | 1.00 | 1.00 | 1.00 | Constant |
| 4 Napoli-Capodichino | 1.00 | 1.00 | 1.00 | Constant |
| 5 Roma-Ciampino-Fiumicino | 1.00 | 1.00 | 1.00 | Constant |
| 6 Venezia-Tessera | 1.00 | 1.00 | 1.00 | Constant |
| 7 Bologna-Borgo Panigale | 0.99 | 1.00 | 0.99 | Increasing |
| 8 Lampedusa | 0.83 | 1.00 | 0.83 | Increasing |
| 9 Perugia | 0.72 | 1.00 | 0.72 | Increasing |
| 10 Grosseto | 0.65 | 1.00 | 0.65 | Increasing |
| 11 Elba | 0.64 | 1.00 | 0.64 | Increasing |
| 12 Bolzano | 0.50 | 1.00 | 0.50 | Increasing |
| 13 Olbia-Costa Smeralda | 0.99 | 0.99 | 0.99 | Increasing |
| 14 Palermo-Punta Raisi | 0.94 | 0.96 | 0.99 | Increasing |
| 15 Genova-Sestri | 0.89 | 0.94 | 0.95 | Increasing |
| 16 Treviso-Sant'Angelo | 0.86 | 0.93 | 0.92 | Increasing |
| 17 LameziaTerme-ReggioCalabria-Crotone | 0.90 | 0.92 | 0.98 | Increasing |
| 18 Firenze-Pisa | 0.91 | 0.91 | 1.00 | Increasing |
| 19 Verona-Brescia | 0.86 | 0.86 | 0.99 | Increasing |
| 20 Trieste-Ronchi dei Legionari | 0.79 | 0.82 | 0.96 | Increasing |
| 21 Torino-Caselle | 0.82 | 0.82 | 0.99 | Increasing |
| 22 Rimini-Miramare | 0.74 | 0.82 | 0.90 | Increasing |
| 23 Cagliari-Elmas | 0.78 | 0.78 | 0.99 | Increasing |
| 24 Bari-Brindisi-Foggia-Taranto | 0.78 | 0.78 | 0.99 | Increasing |
| 25 Alghero-Fertilia | 0.74 | 0.76 | 0.97 | Increasing |
| 26 Pescara | 0.68 | 0.76 | 0.89 | Increasing |
| 27 Cuneo-Levaldigi | 0.50 | 0.66 | 0.75 | Increasing |
| 28 Trapani-Birgi | 0.48 | 0.58 | 0.83 | Increasing |
| 29 Ancona-Falconara | 0.52 | 0.57 | 0.90 | Increasing |
| 30 Parma | 0.23 | 0.48 | 0.48 | Increasing |

Table 3: Application of DEA CCR-I and BCC-I to Italian airports; this table reports the estimated overall efficiency under constant returns ($OTE_{crs}$), the pure technical efficiency under variable returns ($PTE_{vrs}$), the scale efficiency (SE) and the estimated returns to scale (RTS).

| Variables | Min | 1stQ | Median | Mean | 3rd Q | Max |
|---|---|---|---|---|---|---|
| Independent | | | | | | |
| CRS efficiencies | 0.23 | 0.69 | 0.82 | 0.79 | 0.98 | 1.00 |
| VRS efficiencies | 0.48 | 0.79 | 0.93 | 0.88 | 1.00 | 1.00 |
| Dependent | | | | | | |
| Sustainability | 0.00 | 0.00 | 1.00 | 1.73 | 3.00 | 7.00 |
| EBITDA margin | -1.63 | 0.12 | 0.23 | 0.11 | 0.34 | 0.57 |
| Low-cost carrier passengers (%) | 0.00 | 0.41 | 0.61 | 0.58 | 0.73 | 0.99 |
| Ownership (1 = Private) | 0.00 | 0.00 | 0.00 | 0.43 | 1.00 | 1.00 |
| Belongs to group | 0.00 | 0.00 | 0.00 | 0.20 | 0.00 | 1.00 |
| Log pax area ($m^2$) | 5.65 | 8.62 | 9.32 | 9.05 | 9.96 | 11.96 |

Table 4: Summary statistics of the variables employed for the regression analysis.

| Independent Variables | Dependent variable: | |
|---|---|---|
| | OTE (CCR-I) | PTE (BCC-I) |
| | (1) | (2) |
| SUSTAINABILITY | 0.034** | 0.059*** |
| | (0.015) | (0.019) |
| EBITDA | 0.268*** | 0.245*** |
| | (0.033) | (0.039) |
| LCC | 0.027 | -0.157 |
| | (0.076) | (0.100) |
| OWNERSHIP | 0.027 | -0.018 |
| | (0.040) | (0.044) |
| GROUP | 0.006 | 0.037 |
| | (0.052) | (0.061) |
| LOGAREAPAX | 0.040** | -0.086*** |
| | (0.019) | (0.031) |
| Constant | 0.331** | 1.668*** |
| | (0.154) | (0.258) |
| Observations | 30 | 30 |
| Log Likelihood | 18.710 | 9.025 |
| Wald Test (df = 6) | 216.000*** | 120.200*** |
| Pseudo-$R^2$ | 0.779 | 0.668 |

*Note:* *p<0.1; **p<0.05; ***p<0.01

Table 5: Regression results on CCR-I and BCC-I efficiency scores